\documentclass[english,aps,prl,twocolumn,showpacs,amsmath,amssymb,superscriptaddress]{revtex4-1}
\usepackage[T1]{fontenc}
\usepackage[latin9]{inputenc}
\usepackage{bm}
\usepackage{amsmath}
\usepackage{amssymb}
\usepackage{graphicx}

\makeatletter
 \@ifundefined{textcolor}{}
 {%
   \definecolor{BLACK}{gray}{0}
   \definecolor{WHITE}{gray}{1}
   \definecolor{RED}{rgb}{1,0,0}
   \definecolor{GREEN}{rgb}{0,1,0}
   \definecolor{BLUE}{rgb}{0,0,1}
   \definecolor{CYAN}{cmyk}{1,0,0,0}
   \definecolor{MAGENTA}{cmyk}{0,1,0,0}
   \definecolor{YELLOW}{cmyk}{0,0,1,0}
 }

\newcommand{\un}[1]{\,\text{#1}}
\def\bra#1{\langle #1|}
\def\ket#1{|\mbox{$#1$}\rangle}

\def\bracketi#1#2{\langle\mbox{$#1$}|\mbox{$#2$}\rangle}

\makeatother

\usepackage{babel}
\begin{document}

\title{Gate-Control of Spin Precession in Quantum Hall Edge States}

\author{T.~Nakajima}

\email{nakajima@meso.t.u-tokyo.ac.jp}

\altaffiliation[Present address: ]{Department of Applied Physics, University of Tokyo, Hongo, Bunkyo-ku, Tokyo 113-8656, Japan}

\affiliation{Department of Basic Science, University of Tokyo, Komaba, Meguro-ku,
Tokyo 153-8902, Japan}

\author{Kuan-Ting Lin}

\affiliation{Department of Physics, National Tsing Hua University, Hsinchu 30013,
Taiwan}

\author{S.~Komiyama}

\affiliation{Department of Basic Science, University of Tokyo, Komaba, Meguro-ku,
Tokyo 153-8902, Japan}
\begin{abstract}
Electrical control and detection of spin precession are experimentally
demonstrated by using spin-resolved edge states in the integer quantum
Hall regime. Spin precession is triggered at a corner of a biased
metal gate, where electron orbital motion makes a sharp turn leading
to a nonadiabatic change in the effective magnetic field via spin-orbit
interaction. The phase of precession is controlled by the group velocity
of edge-state electrons tuned by gate bias voltage: A spin-FET device
is thus realized by all-electrical means, without invoking ferromagnetic
material. The effect is also interpreted in terms of a Mach-Zehnder-type
spin interferometer.
\end{abstract}


\maketitle
Electrical control of electron spin is a key element for realizing
spin-based quantum information processing in solid-state devices.
Spin-orbit interaction (SOI) has been exploited for manipulating single
spins in quantum dots through alternating electric fields\cite{Nowack:2007uq,Nadj-Perge:2010fk}.
Datta and Das proposed a novel framework for spin manipulation called
spin-FET, in which spin precession of moving electrons in a 1D system
is controlled only by \emph{constant} electric fields\cite{DATTA:1990uq,Koo:2009fk}.
This approach provides a realistic scheme for building an electronic
analogue of photonic quantum information processing\cite{Bertoni:2000fk,Ionicioiu:2001fk,Stace:2004fk,Giovannetti:2008vn,Beenakker:E:K:v:PRL:91:p147901:2003,Samuelsson:S:B:PRL:92:p026805:2004}
in solid-state systems: Quantum logic gates may be implemented in
a coherent circuit by encoding \emph{flying qubits} in the spin degree
of freedom of ballistic electrons.

Edge states in the quantum Hall regime is an ideal 1D electron system
for this purpose because of its large equilibration length\cite{MULLER:1992yt,KOMIYAMA:H:O:M:S:F:PRB:45:p11085-11107:1992}
and long-range quantum mechanical coherence\cite{Ji:C:S:H:M:S:Nature:422:p415-418:2003}.
Moreover, spin-polarized current can be selectively injected and detected
by controlling gate bias conditions\cite{BUTTIKER:PRL:57:p1761-1764:1986,KOMIYAMA:H:S:H:PRB:40:p12566-12569:1989,VANWEES:1989fk}
without resort to ferromagnetic leads assumed in the original proposal
of the spin-FET\cite{DATTA:1990uq}. Indeed, control of spin precession
in edge states has been theoretically discussed\cite{Wang:2004uq,Reynoso:2004fk,Bao:2005fk,Pala:2005oe,grigoryan:165320},
but no experimental realization has been reported so far.

\begin{figure}
\includegraphics[width=8.4cm]{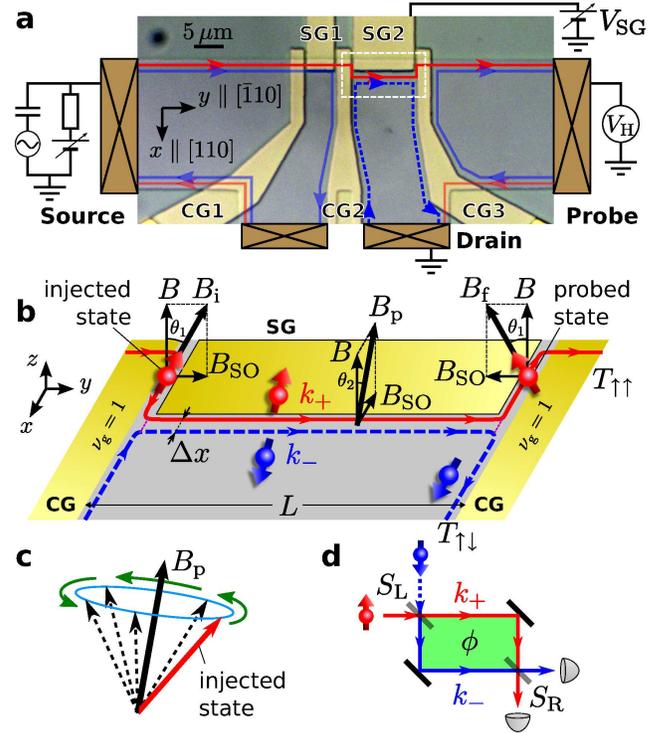}

\caption{(Color online) (a) Optical microscope image of the device studied,
together with a schematic of the experimental setup. Up- and down-spin
edge states in the lowest Landau level are drawn in red and blue lines,
respectively. (b) Schematic view of the region bounded by a white
rectangle in (a). (c) Spin precession during the propagation along
the side gate. (d) Schematic representation of an equivalent Mach-Zehnder-type
spin interferometer.}

\label{Flo:geometry}
\end{figure}

In this Letter, we demonstrate \emph{all-electrical} spin-FET-like
control of spin precession in edge states. The key is a non-uniform
arrangement of SOI-induced magnetic field $\bm{B}{}_{\text{SO}}$
prepared in a tailored geometry of edge states (see Fig.~\ref{Flo:geometry}(b)).
Spin-polarized edge states are quantized in the effective magnetic
field $\bm{B}_{\text{eff}}=\bm{B}{}_{\text{SO}}+\bm{B}$ with $\bm{B}$
being the external magnetic field. Incoming electrons in a spin eigenstate
for $\bm{B}_{\text{eff}}=\bm{B}_{\text{i}}$ are guided to a corner
of a biased metal gate, at which $\bm{B}_{\text{eff}}$ changes to
$\bm{B}_{\text{p}}$ nonadiabatically because $\bm{B}{}_{\text{SO}}$
changes its direction along with the sharp turn of electron drift
motion\cite{KHAETSKII:1992fk,Polyakov:1996uq}. The spin state hence
starts to evolve or precess around the axis of $\bm{B}_{\text{p}}$.
The spin precession continues until the electrons arrive at the opposite
corner of the gate, where the up- and down-spin branches are separated.
The projection of their final state onto the up-spin eigenstate for
$\bm{B}_{\text{eff}}=\bm{B}_{\text{f}}$ is read out by the Hall voltage.
The phase of precession is determined by the Larmor frequency and
the group velocity, which can be experimentally controlled by $\bm{B}$
and the gate bias voltage, respectively.

\begin{figure}
\includegraphics{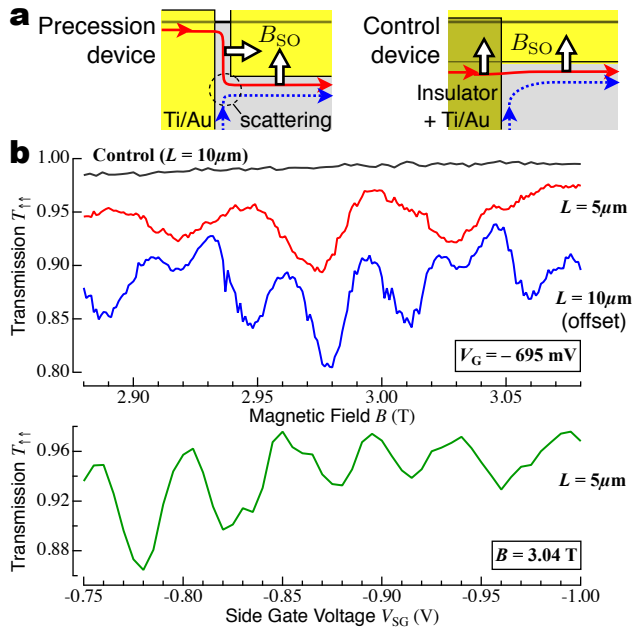}

\caption{(Color online) (a) Schematics of the device in Fig.~\ref{Flo:geometry}
(left) and the control device (right), focusing on the region where
two edge states encounter with each other. (b) $T_{\uparrow\uparrow}$
against $B$ (upper panel) and $V_{\text{SG}}$ (lower panel).}

\label{Flo:oscillation}
\end{figure}

Devices were fabricated in GaAs/AlGaAs heterostructure crystals I
and II%
\footnote{See Supplemental Material for the detailed description of the device
structures and the possible effects of impurities.%
}. Similar experimental results were obtained in the two, and only
the results of crystal I will be described below for reasons of space.
Figure~\ref{Flo:geometry}(a) shows the optical microscope image
of the device with three cross gates (CGs 1-3) and two side gates
(SGs 1 and 2). The filling factor of Landau levels (LLs) in the bulk
region is set to be $\nu=3.7\text{--}2.5$ in magnetic fields of $B=2.7\text{--}4.0\un{T}$
perpendicular to the 2D electron system at $100\un{mK}$. Only the
spin-resolved edge states in the lowest LL are relevant to the experimental
results because the edge states in higher LLs are completely decoupled
and grounded. Two regions defined by CG1, CG2, SG1 (with length $L=5\,\mu\text{m}$)
and by CG2, CG3, SG2 ($L=10\,\mu\text{m}$) are studied separately.
Figure~\ref{Flo:geometry}(a) depicts edge-state trajectories with
$V_{\text{CG1}}=V_{\text{SG1}}=0$ and $V_{\text{CG2}},\, V_{\text{CG3}},\, V_{\text{SG2}}<0$
for studying the region of $L=10\,\mu\text{m}$. Electrons emitted
from the source (drain) contact in the spin-up outer (spin-down inner)
edge states are totally transmitted through (reflected at) CG2, biased
so that $\nu_{\text{g}}=1$ in the region below the gate. The transmitted
electrons change the direction of motion at the left-hand corner of
SG2 ($V_{\text{SG}}<-0.4\un{V}$), start spin precession, and travel
along the boundary of SG2 (see Fig.~\ref{Flo:geometry}(b) and \ref{Flo:oscillation}(a)).
Reaching the right-hand corner of SG2, the electrons are partly transmitted
(reflected) with probability $T_{\uparrow\uparrow}$ ( $T_{\uparrow\downarrow}=1-T_{\uparrow\uparrow}$),
which is experimentally probed by $T_{\uparrow\uparrow}=V_{\text{H}}/V_{\text{S}}$
with $V_{\text{H}}$ and $V_{\text{S}}$ being the Hall and the source
voltages\cite{BUTTIKER:PRL:57:p1761-1764:1986,KOMIYAMA:H:S:H:PRB:40:p12566-12569:1989,VANWEES:1989fk}.
The voltage $V_{\text{H}}$ is studied via a lock-in technique by
superposing an ac component ($V_{\text{ac}}=12.9\,\mu\text{V}$ at
$130\un{Hz}$) to a dc background ($V_{\text{dc}}=-103\,\mu\text{V}$)
in the source voltage ($V_{\text{S}}=V_{\text{ac}}+V_{\text{dc}}$)\cite{Note1}.

The spin precession manifests itself as distinct oscillations in $T_{\uparrow\uparrow}$
against $B$, as displayed in the upper panel of Fig.~\ref{Flo:oscillation}(b).
The oscillation is visible both in the $10\,\mu\text{m}$- and the
$5\,\mu\text{m}$-long regions\cite{Note1}, where the oscillation
period for the latter is approximately twice as large as that for
the former. The oscillation shows up also as a function of $V_{\text{SG}}$
at a fixed $B$ (lower panel of Fig.~\ref{Flo:oscillation}(b)). 

To make explicit the essential role of the side-gate corners, a control
experiment was made on another device shown in the right panel of
Fig.~\ref{Flo:oscillation}(a), where the sharp turn of electron
motion is prevented at the left end of the SG region. The device is
prepared with CGs and SG overlapping with each other by introducing
an insulation layer in between\cite{Note1}. The oscillatory behavior
is completely absent in the control experiment, as exemplified by
the topmost line in Fig.~\ref{Flo:oscillation}(b). This provides
compelling evidence that the oscillation of $T_{\uparrow\uparrow}$
is triggered by the nonadiabatic change of $\bm{B}{}_{\text{SO}}$
induced at the left corner of SG. In addition, the values of $T_{\uparrow\uparrow}$
in the control device are close to unity ($T_{\uparrow\uparrow}\approx0.99$),
being almost independent of $B$ (and of $V_{\text{\text{SG}}}$ though
not shown here). The nearly perfect transmission indicates macroscopic
equilibration lengths\cite{MULLER:1992yt,KOMIYAMA:H:O:M:S:F:PRB:45:p11085-11107:1992},
$\ell_{\text{eq}}\sim300\,\mu\text{m}\gg L$ in the present experiment,
suggesting that the electrons are practically free from impurity-induced
spin-flip scattering.

\begin{figure}
\includegraphics{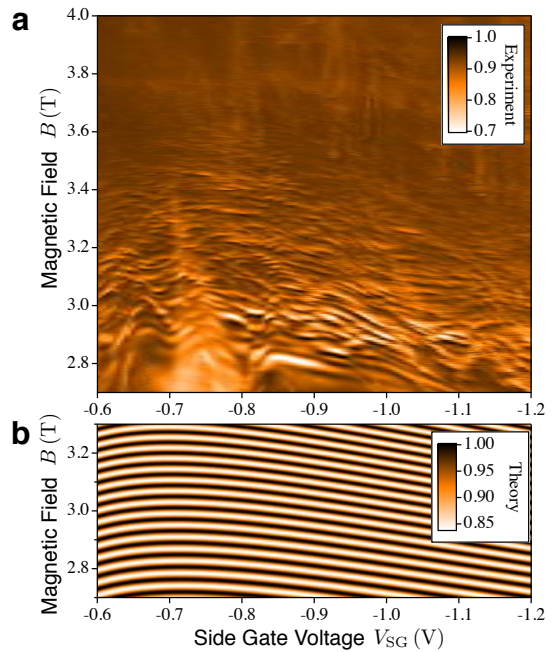}

\caption{(Color online) Two-dimensional plot of $T_{\uparrow\uparrow}$ on
the$B$-$V_{\text{G}}$ plane. (a) Experimental results obtained in
the device of $L=10\,\mu\text{{m}}$. (b) Theoretical results derived
from eq.~\eqref{eq:transmission} with $g=-2.9$, $C_{0}=0.92$,
$C_{1}=0.08$ and $\phi_{0}=0$.}

\label{Flo:BvsVg}
\end{figure}

Figure~\ref{Flo:BvsVg}(a) displays a two-dimensional plot of $T_{\uparrow\uparrow}(B,\, V_{\text{SG}})$
for $L=10\,\mu\text{m}$. The oscillatory pattern is visible in a
wide range of $B$ and $V_{\text{SG}}$ such that $B=2.7$--$3.5\un{T}$
and $V_{\text{SG}}=-0.6$--$-1.2\un{V}$. Although the pattern is
disturbed to some extent by irregular structures, general trend of
oscillation is evident and the overall feature is distinctly different
from the well-known irregular structure due to impurity-induced scattering\cite{Note1}.

For quantitative interpretation, we take into account two components
of SOI in GaAs/AlGaAs heterostructures; namely, Rashba and Dresselhaus
terms\cite{Miller:2003qz}. The two terms are of comparable strengths,
and the Hamiltonian of each is given by $H_{\text{R}}=\alpha(p_{y}\sigma_{x}-p_{x}\sigma_{y})/\hbar$
and $H_{\text{D}}=\beta(p_{y}\sigma_{x}+p_{x}\sigma_{y})/\hbar$ in
the coordinate system of $x\parallel[110]$, $y\parallel[\bar{1}10]$,
and $z\parallel[001]$. Here $p_{i}$ is the electron momentum and
$\sigma_{i}$ is the Pauli matrix ($i=x$, $y$, or $z$). These terms
yield the effective magnetic field $\bm{B}_{\text{SO}}=(2\alpha^{\prime}p_{y}/(g\mu_{\text{B}}\hbar),\,2\beta^{\prime}p_{x}/(g\mu_{\text{B}}\hbar),\,0)$,
with $g$ the effective $g$-factor ($g<0$), $\mu_{\text{B}}$ the
Bohr magneton, $\alpha^{\prime}=\alpha+\beta$, and $\beta^{\prime}=-\alpha+\beta$
($|\alpha^{\prime}|\gg|\beta^{\prime}|$, $\beta^{\prime}<0$). This
representation proves to be valid for edge states when $\bm{p}$ is
replaced with $m\bm{v}_{g}$, where $m$ and $\bm{v}_{g}$ are the
effective mass and the group velocity of edge-state electrons\cite{KHAETSKII:1992fk,Polyakov:1996uq}.
In the device studied (Fig.~\ref{Flo:geometry}(b)), $\bm{B}_{\text{SO}}$
changes from $(0,\, B_{\text{SOi}},\,0)$ to $(B_{\text{SOp}},\,0,\,0)$
at the left-hand corner of SG where $B_{\text{SOi}}=2\beta^{\prime}m|\bm{v}_{g}|/(g\mu_{\text{B}}\hbar)$
and $B_{\text{SOp}}=2\alpha^{\prime}m|\bm{v}_{g}|/(g\mu_{\text{B}}\hbar)$.
The spinor of the incoming electrons is given by $\ket{i}=R_{x}(-\theta_{1})\ket{\uparrow}$
where $\ket{\uparrow}$ is the eigenstate of $\sigma_{z}$, $R_{i}(\theta)$
is the rotation operator about the $i$-axis, and $\theta_{1}=\arctan(B_{\text{SOi}}/B)$.
After turning the corner, the up- and the down-spin branches along
the SG boundary are similarly given by $\ket{+}=R_{y}(-\theta_{2})\ket{\uparrow}$
and $\ket{-}=R_{y}(-\theta_{2})\ket{\downarrow}$ with $\theta_{2}=\arctan(B_{\text{SOp}}/B)$.
Hence, the initial state $\ket{i}$ becomes a superposition $a\ket{+}+b\ket{-}$
at the corner where the direction of $\bm{v}_{g}$ changes nonadiabatically%
\footnote{In the limit of sharp potential edge at the gate corner, $a$ and
$b$ are determined so that $\ket{i}=a\ket{+}+b\ket{-}$. In this
case, $C_{1}\sim0.08$ is derived for $B=3\un{T}$. In the opposite
limit of smooth corner, $\ket{i}$ would adiabatically changes to
$\ket{+}$; i.e., $C_{1}=0$ or no inter-edge state scattering takes
place ($S_{\text{L}}=S_{\text{R}}=I$). The values of $a$ and $b$
are also influenced by random potential fluctuations in real devices.%
}. In other words, coherent inter-edge state scattering is caused by
the local electric field at the gate corner via the SOI\cite{KHAETSKII:1992fk,Polyakov:1996uq}.
In the control device, $\bm{B}_{\text{SO}}=(B_{\text{SO}},\,0,\,0)$
is invariable with $B_{\text{SO}}=2\alpha^{\prime}m|\bm{v}_{g}|/(g\mu_{\text{B}}\hbar)$
along SG (Fig.~\ref{Flo:oscillation}(a)) so that the spin state
is kept unchanged.

The up- and down-spin branches, $\ket{+}$ and $\ket{-}$, are associated
with quasi-one-dimensional electron waves that propagate along the
SG boundary (in the $y$-direction) with different wave numbers $k_{\pm}=-x_{\pm}/\ell_{\text{B}}^{2}$,
where $\ell_{\text{B}}=\sqrt{\hbar/(eB)}\sim15\un{nm}$ is the magnetic
length. The wave functions of the edge states are given by $\varphi_{k\pm}(x,y)=\frac{1}{\sqrt{L}}e^{ik_{\pm}y}\chi_{0}(x-x_{\pm})\ket{\pm}$\cite{Yoshioka:2002fk},
where $\chi_{0}(x-x_{\pm})\propto\exp(-(x-x_{\pm})^{2}/2\ell_{\text{B}}^{2})$
describes the $x$-profile about the center coordinates $x=x_{\pm}$.
During the propagation, the state evolves as $\ket{\psi(y)}=ae^{ik_{+}y}\ket{+}+be^{ik_{-}y}\ket{-}$
(here the $x$-profile $\chi_{0}$ is ignored). This yields the spin
precession $\bra{\psi}\sigma_{y}\ket{\psi}\propto\cos(\phi(y)+\phi_{0})$
with the phase $\phi(y)=(k_{+}-k_{-})y$ as illustrated in Fig.~\ref{Flo:geometry}(c).
Physical implication of the phase is made explicit by rewriting $\phi(y)$
as 
\begin{equation}
\phi(y)=\omega_{\text{L}}y/|\bm{v}_{g}|\,,\label{eq:precession}
\end{equation}
where $\omega_{\text{L}}=|g\mu_{\text{B}}B|/\hbar$ is the Larmor
frequency and $|\bm{v}_{g}|=(1/\hbar)\left|g\mu_{\text{B}}B/\Delta k\right|$
with $\Delta k=k_{-}-k_{+}$. Equation~\eqref{eq:precession} indicates
that the phase of precession is experimentally controlled by $B$
through $\omega_{\text{L}}$ as well as by $V_{\text{SG}}$ through
$|\bm{v}_{g}|$.

Similar coherent scattering takes place at the right-hand corner ($y=L$).
The total transmission probability is given by the projection of $\ket{\psi(L)}$
onto the ``probed state'', $\ket{f}=R_{x}(+\theta_{1})\ket{\uparrow}$,
in the up-spin branch of outgoing edge states: 
\begin{eqnarray}
T_{\uparrow\uparrow}=|\bracketi{f}{\psi(L)}|^{2} & = & \left|\bra{\uparrow}S_{\text{R}}\left(\begin{array}{cc}
e^{ik_{+}y} & 0\\
0 & e^{ik_{-}y}
\end{array}\right)S_{\text{L}}\ket{\uparrow}\right|^{2}\nonumber \\
 & = & C_{0}+C_{1}\cos(\phi(L)+\phi_{0})\,.\label{eq:transmission}
\end{eqnarray}
The offset $C_{0}$, the amplitude $C_{1}$, and the phase offset
$\phi_{0}$ are constants determined by the scattering matrices $S_{\text{L}}$
and $S_{\text{R}}$ at the side-gate corners\cite{Note2}. The phase
of precession at $y=L$ is hence detected from the oscillation in
$T_{\uparrow\uparrow}$.

By noting $\Delta k=-\Delta x/\ell_{\text{B}}^{2}$ with $\Delta x=x_{-}-x_{+}$,
we can also express $\phi$ as the Aharonov-Bohm phase 
\begin{equation}
\phi(L)=2\pi BL\Delta x/(h/e)\,,\label{eq:ABphase}
\end{equation}
determined by the number of magnetic flux quanta threading through
the narrow stripe enclosed by the co-propagating edge states. This
leads us to the equivalent interpretation in terms of the Mach-Zehnder
type interferometer (see Fig.~\ref{Flo:geometry}(d)\cite{Ji:C:S:H:M:S:Nature:422:p415-418:2003}),
where $T_{\uparrow\uparrow}$ oscillates as a result of the interference
of two paths between ``beam splitters'' represented by the scattering
matrices $S_{\text{L}}$ and $S_{\text{R}}$. This system thus serves
as a spin filtering device based on the two-path interference\cite{Lopez:2010fk}.

\begin{figure}
\includegraphics{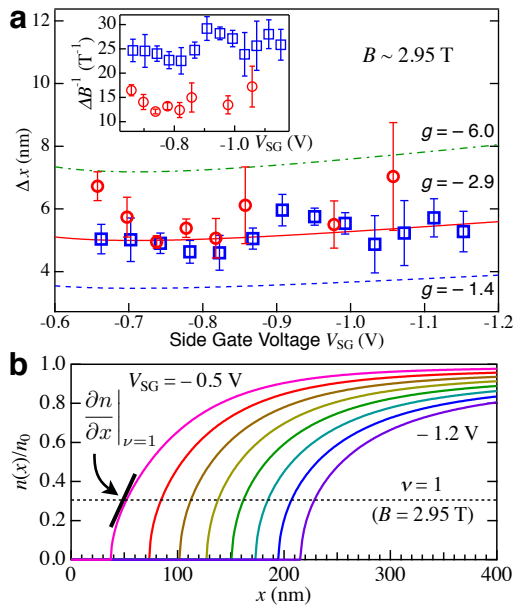}

\caption{(Color online) (a) Inter-edge state separation $\Delta x$ against
$V_{\text{SG}}$ for $L=5\,\mu\text{{m}}$ (red circles) and $L=10\,\mu\text{{m}}$
(blue squares), derived from $\Delta B^{-1}$ values in the inset.
Dashed, solid, and dash-dotted lines represent theoretical values
derived for $g=-1.4$, $-2.9$, and $-6.0$. The inset shows experimental
values of $\Delta B^{-1}$ obtained from the oscillation in a range
of $B=2.85$-$3.05\un{T}$. (b) Theoretically expected electron density
distribution $n(x)$ for $V_{\text{SG}}=-0.5,\,-0.6,\,\cdots,\,-1.2\un{V}$\cite{LARKIN:1995yb}.
Edge states are located at positions where $n(x)$ corresponds to
$\nu=1$, as indicated by the horizontal dashed line for $B\sim2.95\un{T}$.}

\label{Flo:frequency}
\end{figure}

Oscillations in $T_{\uparrow\uparrow}(B)$ for different values of
$V_{\text{SG}}$ are analyzed with fast Fourier transform in a range
of $B=2.85\text{--}3.05\un{T}$\cite{Note1}, and the derived frequencies,
$\Delta B^{-1}$, are plotted against $V_{\text{SG}}$ in the inset
of Fig.~\ref{Flo:frequency}(a). This confirms that $\Delta B^{-1}$
for $L=10\,\mu\text{{m}}$ is nearly twice as large as that for $L=5\,\mu\text{{m}}$.
The values of $\Delta B^{-1}$ are translated into those of $\Delta x$
through eq.~\eqref{eq:ABphase} as shown in Fig.~\ref{Flo:frequency}(a),
which suggests that the spin-split inter-edge state distance is $\Delta x\sim5\un{nm}$
($\ll\ell_{\text{B}}$).

These values of $\Delta x$ are consistent with a theoretical expression
$\Delta x\approx\sqrt{8|g\mu_{\text{B}}B|\varepsilon\varepsilon_{0}/(\pi e^{2}\left.dn/dx\right|_{\nu=1})}$
with $\varepsilon=13$\cite{CHKLOVSKII:S:G:PRB:46:p4026-4034:1992}.
Here the electron density profile near the edge $n(x)$ is derived
from an electrostatic model\cite{LARKIN:1995yb} and is plotted in
Fig.~\ref{Flo:frequency}(b) for different values of $V_{\text{SG}}$.
Theoretical values of $\Delta x$ are indicated with different lines
in Fig.~\ref{Flo:frequency}(a) for $g=-1.4$, $-2.9$, and $-6.0$.
Reasonable agreement with the experiment is obtained with $g=-2.9\pm0.1$,
which is consistent with the exchange-enhanced $g$-factor at $\nu=1$,
estimated to be $g=-2\text{--}-7$ in earlier experiments\cite{NICHOLAS:1988fk,USHER:1990fk}.
(Note that the enhanced $g$-factor includes the effect of SOI\cite{MULLER:1992yt}.)

Closer look at the experimental data in Figs.~\ref{Flo:BvsVg}(a)
and \ref{Flo:frequency}(a) indicates that $\Delta B$ tends to decrease
with increasing $B$ and decreasing $V_{\text{SG}}$ ($<-0.7\un{V}$).
These trends are also explained with a theoretically expected variation
of $\Delta x$ (or $\bm{\nu}_{g}$): For instance, the profile of
$n(x)$ in Fig.~\ref{Flo:frequency}(b) shows that the confining
potential becomes less steep and $\Delta x$ increases as the edge
states are pushed further away from the SG boundary with decreasing
$V_{\text{SG}}$\cite{LARKIN:1995yb}. It is this dependence that
gives rise to the oscillation against $V_{\text{SG}}$ shown in the
lower panel of Fig.~\ref{Flo:oscillation}(b).

Displayed in Fig~\ref{Flo:BvsVg}(b) is a 2D plot of the theoretically
predicted oscillatory pattern of $T_{\uparrow\uparrow}$ in a $B$-$V_{\text{SG}}$
plane, which is derived from eq.~\eqref{eq:transmission} with $g=-2.9$,
$C_{0}=0.92$, $C_{1}=0.08$, and $\phi_{0}=0$\cite{Note2}. Equiphase
lines substantially reproduce the overall feature of experimental
oscillations in Fig.~\ref{Flo:BvsVg}(a). (The equiphase lines take
broad maxima at $V_{\text{SG}}\approx-0.7\un{V}$ giving opposite
slope on the side of $V_{\text{SG}}>-0.7\un{V}$. This is consistent
with the experiment and arises from the fact that the confining potential
in Fig.~\ref{Flo:frequency}(b) gets steeper as $V_{\text{SG}}$
decreases for $V_{\text{SG}}>-0.7\un{V}$\cite{LARKIN:1995yb}.)

The experimental pattern in Fig.~\ref{Flo:BvsVg}(a) suffers from
irregular distortion. This is probably because long-range random potential
affects the landscape of edge confining potential, giving rise to
local fluctuations of $\Delta x$ or $\bm{v}_{g}$\cite{Note1}. The
area enclosed by the co-propagating trajectories of edge states, $L\Delta x$,
or the phase $\phi(L)$ of spin precession, hence fluctuates as $B$
or $V_{\text{SG}}$ varies. Such long-range potential fluctuation,
however, does not cause inter-edge state scattering.

We note in Fig.~\ref{Flo:BvsVg}(a) that $T_{\uparrow\uparrow}$
approaches unity and the amplitude of oscillation significantly diminishes
as $B$ increases beyond $\sim3.4\un{T}$. This implies that the scattering
probability at the SG corners decreases at higher $B$. This is readily
interpreted by noting that the potential at the side-gate corners
becomes smoother due to screening by a wider compressible region of
$0<\nu<1$ as $B$ increases. The edge-state electrons in this regime
adiabatically change their direction of motion in a smoothly curved
potential without undergoing inter-edge state scattering\cite{Note2}.
The decrease of wave function overlap with increasing Zeeman splitting
or decreasing $\ell_{\text{B}}$ may also suppress the scattering,
but this effect was confirmed to be insignificant in our separate
experiments with tilted magnetic fields.

The amplitude of oscillation corresponds to $C_{1}\sim0.04$ at $B\sim3\un{T}$
in the experiment, which is smaller than but on the same order of
the predicted value in the nonadiabatic limit\cite{Note2}. This value
is similar between the regions of $L=5\,\mu\text{m}$ and $10\,\mu\text{m}$
(Fig.~\ref{Flo:oscillation}(b)) and confirmed to be independent
of temperature up to $\sim160\un{mK}$. These findings suggest that
the visibility of the oscillation is restricted by $S_{\text{L}}$
and $S_{\text{R}}$ in eq.~\eqref{eq:transmission} and the spin
coherence length well exceeds $10\,\mu\text{m}$ at $100\un{mK}$,
which is noticeably larger than that of the orbital motion\cite{Ji:C:S:H:M:S:Nature:422:p415-418:2003}.
All-electrical coherent control of flying spins demonstrated in this
work suggests that the edge states is a promising candidate for implementing
flying spin qubits.

In conclusion, electrical control of spin precession in edge states
has been demonstrated by converting gate-bias-induced static electric
fields into spatially varying effective magnetic fields through the
SOI. The phase of spin precession is controlled by\textbf{ $B$} (through
the Larmor frequency) or by $V_{\text{SG}}$ (through the group velocity
of electrons). We suggest an equivalent interpretation in terms of
a Mach-Zehnder-type spin interferometer, in which the magnetic flux
threading the area bounded by co-propagating spin-split edge states
is controlled either by $B$ or by $L\Delta x$. A ferromagnet-free
spin-FET device has thus been experimentally demonstrated.

\bibliographystyle{apsrev4-1}

\end{document}